\documentclass[12pt,a4paper]{iopart}

\usepackage{iopams}
\usepackage{graphicx,subfigure}

\eqnobysec

\newcommand{\msc}[1]{\vspace{10pt}
     \begin{indented}
     \item[]\rm Mathematics Subject Classification: #1\par
     \end{indented}}

\newcommand{\mydate}[1]{\vspace{5pt}
        \begin{indented}
        \item[]\rm #1\par
        \end{indented}}

\newcommand{\UN}{{\rm U}(N)}

\newcommand{\abs}[1]{\left | #1 \right |}
\newcommand{\rpart}{\mathop{\mathrm{Re}}\nolimits}
\newcommand{\ipart}{\mathop{\mathrm{Im}}\nolimits}
\newcommand{\Ip}{\mathop{\mathrm{Ip}}\nolimits}

\begin{document}
\title{Random matrix theory and the zeros of $\zeta\,'(s)$ }

\author{Francesco Mezzadri}

\address{School of Mathematics, University of Bristol,%
 University Walk, Bristol, BS8 1TW, UK}

\ead{f.mezzadri@bristol.ac.uk}

\mydate{5 November 2002}

\begin{abstract}
We study the density of the roots of the derivative of the
characteristic polynomial $Z(U,z)$ of an $N \times N$ random
unitary matrix with distribution given by Haar measure on the
unitary group. Based on previous random matrix theory models of
the Riemann zeta function $\zeta(s)$, this is expected to be an
accurate description for the horizontal distribution of the zeros
of $\zeta\,'(s)$ to the right of the critical line. We show that
as $N \rightarrow \infty$ the fraction of the roots of $Z\,'(U,z)$
that lie in the region $1-x/(N-1) \le \abs{z} < 1$ tends to a
limit function. We derive asymptotic expressions for this function
in the limits $x \rightarrow \infty$ and $x \rightarrow 0$ and
compare them with numerical experiments.
\end{abstract}
\msc{15A52, 11M99}

\section{Introduction}
\label{int}
We study the density of the roots of
\begin{equation*}
\label{dcp}
 Z\,'(U,z) = \frac{\rmd}{\rmd z} \det(Iz -
U) = \frac{\rmd}{\rmd z} \prod_{j=1}^N \left(z - \rme^{\rmi
\theta_j}\right) , \quad z \in \mathbb{C},
\end{equation*}
where $U$ is a random $N \times N$ unitary matrix, with respect to
the  circular unitary ensemble (CUE) of random matrix theory
(RMT). Our main motivation is to investigate the horizontal
distribution of the zeros of the derivative of the Riemann zeta
function.

The zeta function is defined by
\[
\zeta(s)=\sum_{n=1}^{\infty}\frac{1}{n^s}, \quad \sigma=\rpart(s)
> 1,
\]
and has an analytic continuation in the rest of the complex plane
except for a simple pole at $s=1$. There are infinitely many {\it
non-trivial} solutions to the equation $\zeta(s)=0$ in the strip
$0 < \sigma < 1$; the Riemann hypothesis (RH) states that they all
lie on the {\it critical line} $\sigma=1/2$.  The interest in the
horizontal distribution of the zeros of $\zeta\,'(s)$ is motivated
by its connection with RH. In 1934 Speiser~\cite{Spe34} showed
that RH is equivalent to $\zeta\,'(s)$ having no zeros in the
region $0 < \sigma < 1/2$. Furthermore, up to now the most
efficient ways of computing  the fraction the of zeros of the
Riemann zeta function on the the critical line are based on what
is known as {\it Levinson's method}~\cite{Con89}; it turns out
that the zeros of $\zeta\,'(s)$ close to the critical line have a
significant effect on the efficiency of this
technique~\cite{CG90}, therefore it is important to know how they
are distributed. Levinson and Montgomery~\cite{LM74} proved a
quantitative refinement of Speiser's theorem, namely that
$\zeta(s)$ and $\zeta\,'(s)$ have essentially the same number of
zeros to the left of $\sigma = 1/2$, and showed that as $T
\rightarrow \infty$, where $T$ is the height on the critical line,
a positive proportion of the zeros of $\zeta\,'(s)$ are in the
region
\[
\sigma < \frac12 + (1 + \epsilon) \frac{\log \log T}{\log T},
\quad \epsilon > 0.
\]
Subsequent improvements of Levinson and Montgomery's results,
first by Conrey and Ghosh~\cite{CG90}, then by Guo~\cite{Guo96},
Soundararajan~\cite{Sou98} and recently by Zhang~\cite{Zha01} have
established that a typical zero of $\zeta\,'(s)$ tends to be much
closer to the critical line and that conditionally on RH a
positive proportion lie in the region
\[
\sigma < \frac12 + \frac{C}{\log T},
\]
for some positive constant $C$.  Their distribution, however, is
still unknown. Other results on the zeros of $\zeta\,'(s)$ can be
found in~\cite{Ber70}.

Over the past thirty years, overwhelming evidence has been
accumulated which suggests that the local correlations of the
non-trivial zeros of $\zeta(s)$ coincide, as $T \rightarrow
\infty$, with those of the eigenvalues of hermitian matrices of
large dimensions from the Gaussian unitary ensemble
(GUE)~\cite{Ber86}. As $N \rightarrow \infty$, the GUE statistics
are in turn the same as those of the phases of the eigenvalues of
$N \times N$ unitary matrices, on the scale of their mean distance
$2\pi/N$, averaged over the CUE ensemble. More recently, however,
it was realized that RMT not only describes with high accuracy the
distribution of the Riemann zeros, but that it also provides
techniques to make predictions and computations about the Riemann
zeta function and certain classes of {\it L}-functions that
previous methods had not been able to tackle. This started with
the work of Keating and Snaith~\cite{KS00a} on moments of the
Riemann zeta function and other {\it L}-functions. Their key
observation was that the locally determined statistical properties
of $\zeta(s)$ high up the critical line can be modelled by
characteristic polynomials $Z(U,z)$ of random unitary matrices
$U$. In this model the two asymptotic parameters, $T$ for
$\zeta(s)$ and $N$ for $U$, are compared by setting the densities
of the zeros of $\zeta(s)$ and of the eigenvalues of $U$ equal,
i.e.
\[
N=\log \frac{T}{2\pi}.
\]
 This approach has since been extremely successful~\cite{CFKRS02}.

Following the same ideas, in this paper we suggest that the
density $\rho(z)$ of the roots of
 $Z\,'(U,z)$ will accurately describe the distribution
of the zeros of $\zeta\,'(s)$. A classical theorem in complex
analysis states that if $p(z)$ is a polynomial, then the roots of
$p\,'(z)$ that are not roots of $p(z)$ lie all in the interior or
on the boundary of the smallest convex polygon containing the
zeros of $p(z)$ (see, e.g.,~\cite{PS72}). Therefore, since the
eigenvalues of a unitary matrix have modulus one, the solutions of
the equation $Z\,'(U,z)=0$ that are not zeros of $Z(U,z)$ are all
inside the unit circle. If $s = 1/2 + \rmi t + u$, $t \in
\mathbb{R}$, denotes the point at which $\zeta\,'(s)$ is
evaluated, then the region of $\mathbb{C}$ to the right of the
critical line is mapped inside the unit circle by the conformal
mapping $z = \rme^{-u}$. Thus, the radial density
\[
\int_0^{2\pi} \abs{z} \rho(z)\; \rmd \phi
\]
becomes the analogue of the horizontal distribution of the zeros
of $\zeta\,'(s)$ to the right of the line $\sigma=1/2$. However,
instead of $\rho(z)$, it turns out to be more convenient to
consider $\Ip(x)$, the fraction of the roots in the annulus $1
-x/(N-1) \le \abs{z} < 1$, where $x$ is the {\it scaled distance}
from the unit circle. Our main results concern the asymptotics of
$\Ip(x)$. We show that as $N$ increases the roots of $Z\,'(U,z)$
approach the unit circle, and $\Ip(x)$ tends to a function
independent of $N$. Furthermore, we obtain the following
asymptotics as $N \rightarrow \infty$:
\begin{eqnarray*}
\Ip(x) \sim 1 - 1/x, \quad x \rightarrow \infty, \quad x={\rm o}(N), \\
 \Ip(x) = \frac{8}{9\pi}\; x^{3/2} - \frac{64}{225\pi}\;x^{5/2} +
\frac{128}{2205\pi}\;x^{7/2} + \Or(x^4).
\end{eqnarray*}
These formulae are then tested numerically.

The paper is organized as follows: in section~\ref{zUp} we
introduce the mathematical problem and describe the main
properties of the roots of $Z\,'(U,z)$; in section~\ref{asrho} the
asymptotics of $\Ip(x)$ as $x \rightarrow \infty$ are computed;
using heurstic arguments, in section~\ref{smxas}, $\Ip(x)$ is
derived in the limit $x \rightarrow 0$; section~\ref{concl}
concludes the paper with final remarks.

\section{The distribution of the roots of $Z\,'(U,z)$}
\label{zUp}

The CUE ensemble of RMT is defined as the space $\UN$ of $N \times
N$ unitary matrices endowed with a probability measure
$\rmd\mu(U)$ invariant under any inner automorphism
\begin{equation*}
\label{aut}
 U \mapsto VUW,\quad U,V,W \in \UN.
\end{equation*}
In other words, $\rmd\mu(U)$ must be invariant under left and
right multiplication by elements of $\UN$, so that each matrix in
the ensemble is equally weighted. There exists a unique measure on
the unitary group $\UN$ with this property, known as {\it Haar
measure}. The infinitesimal volume element of the CUE ensemble
occupied by those matrices whose eigenvalues have phases lying
between $\btheta=(\theta_1,\theta_2,\ldots, \theta_N)$ and
$\btheta + \rmd^N \btheta$ is given by
\begin{equation}
\label{haar}
\Omega(N) \Delta^2(\btheta) \rmd^N \btheta,
\end{equation}
 where
\[
\Delta(\btheta)=\prod_{1\, \le j\, <\, k\, \le \, N}
\abs{\rme^{\rmi \theta_j} - \rme^{\rmi \theta_k}} \quad {\rm and}
\quad \Omega(N)=\frac{1}{(2\pi)^N N!}.
\]
Our goal in this paper is to study the density of the roots of
$Z\,'(U,z)$, where $U$ is a random unitary matrix with
distribution given by~\eref{haar}. The analogous problem for
the Ginibre ensemble has been studied by Dennis and
Hannay~\cite{DH02}.

\begin{figure}
\centering \subfigure[$N=20$]{
\includegraphics[width=3.0in]{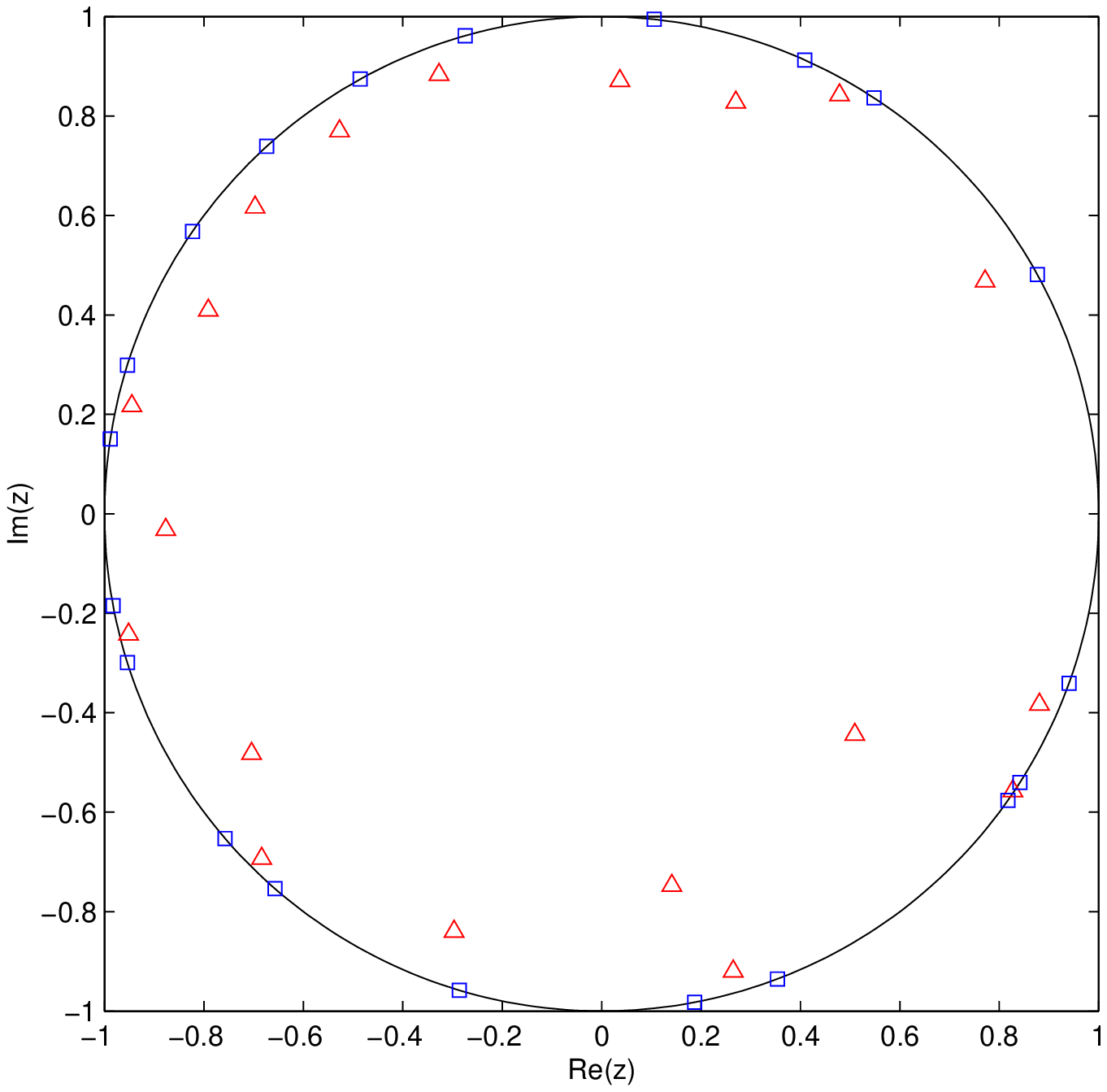}}
 \subfigure[$N=50$]{
  \includegraphics[width=3.0in]{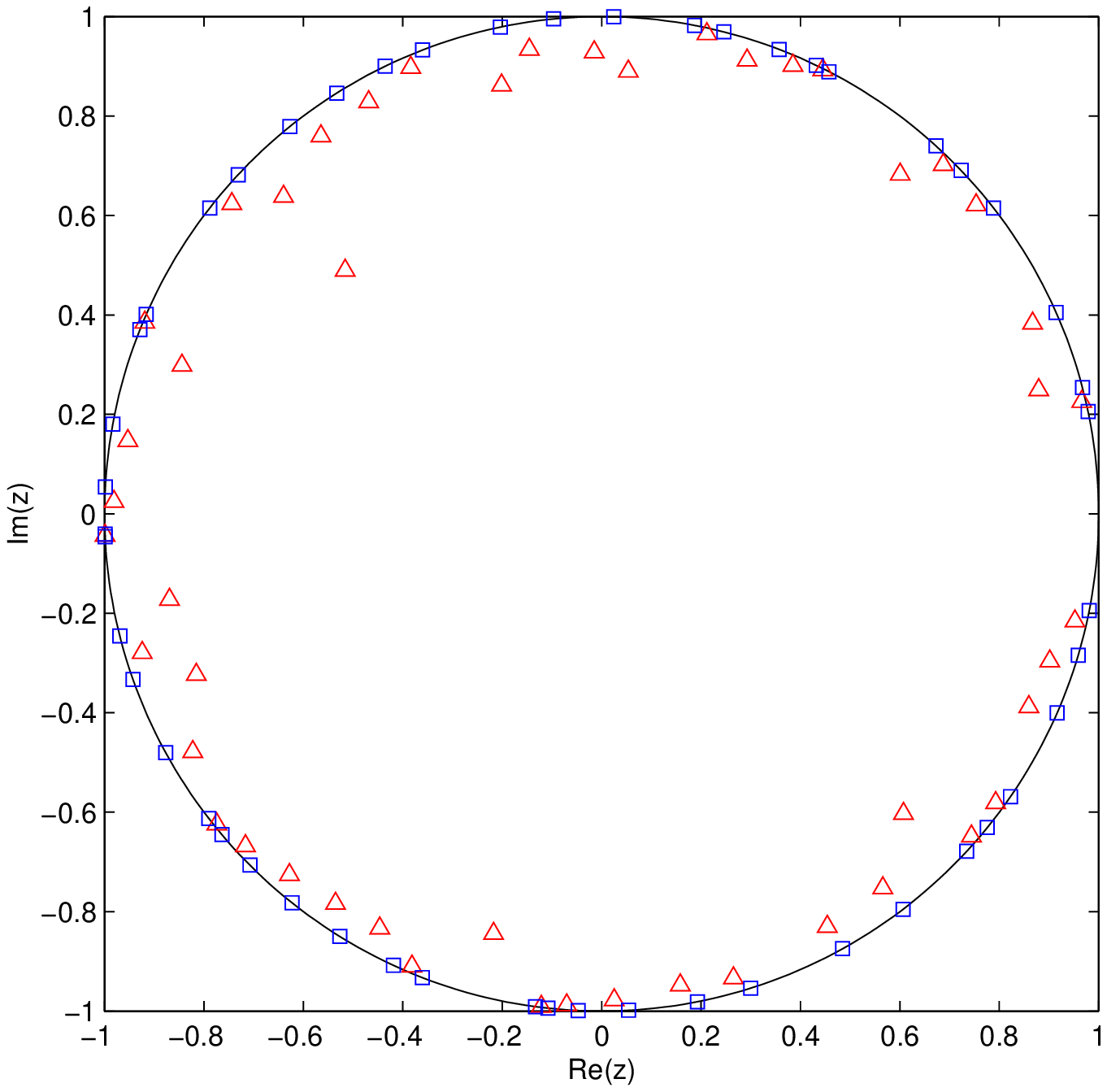}}
      \caption{Zeros of characteristic polynomials of random unitary
      matrices (\opensquare) and of their derivatives (\opentriangle).}
      \label{fig1}
\end{figure}
In figure~\ref{fig1} are plotted the zeros of the characteristic
polynomials $Z(U,z)$ and of their derivatives of two unitary
matrices taken at random with respect to Haar measure for $N=20$
and $N=50$.  Such matrices can be easily generated numerically by
taking complex matrices whose entries are independent complex
random numbers with Gaussian distribution, and then by applying
Gram-Schmidt orthogonalization to the rows or columns (see,
e.g.,~\cite{Rei97}). There are a few qualitative features that can
be immediately observed. Firstly, since the
distribution~\eref{haar} is translation invariant on the unit
circle, the density of the roots of $Z\,'(U,z)$ depends only on
the distance from the origin. Secondly, as mentioned in the
introduction, the roots of $Z\,'(U,z)$ are all inside the unit
circle.  This property can be easily understood with the following
argument. Let $z_1,z_2,\ldots,z_N$ be $N$ complex numbers; if they
all are on the same side of a straight line passing through the
origin, then
\begin{equation}
\label{sums}
 z_1 + z_2 + \cdots + z_N \neq 0 \quad {\rm and} \quad
\frac{1}{z_1} + \frac{1}{z_2} + \cdots + \frac{1}{z_N} \neq 0.
\end{equation}
 Let $\{z_j\}_{j=1}^N$ be the roots of a polynomial $p(z)$ and
$z$ a point outside the smallest convex polygon containing
$\{z_j\}_{j=1}^N$. Now, consider a straight line passing through
$z$ and lying outside such a polygon. Because of
equation~\eref{sums}, the logarithmic derivative
\[
\frac{p\,'(z)}{p(z)} = \frac{1}{z- z_1} + \frac{1}{z-z_2} + \cdots
+ \frac{1}{z-z_N}
\]
cannot vanish. There are two others less obvious features that
figure~\ref{fig1} reveals and that become more apparent as $N$
increases: firstly, the roots of $Z\,'(U,z)$ concentrate in a
small region in proximity of the the unit circle; secondly, given
two consecutive zeros of $Z(U,z)$, say $\rme^{\rmi \theta_j}$ and
$\rme^{\rmi \theta_{j+1}}$, close to each other, there often
appears to be a root of $Z\,'(U,z)$ near the midpoint
\begin{equation}
\label{mpoin} \frac{\rme^{\rmi \theta_j} + \rme^{\rmi
\theta_{j+1}}}{2}.
\end{equation}
In the following sections we shall give quantitative
interpretations of these properties.

Let us set $u = \rpart{z}$ and $v=\ipart{z}$; furthermore, denote
by $\{\lambda_j(\btheta) \}_{j=1}^{N-1}$ the set of roots of
$Z\,'(U,z)$ and consider the linear functional
\[
F[\tau]_{\lambda_j(\btheta)} = \int_{\mathbb{C}} \tau(z)
\bdelta\left(z - \lambda_j(\btheta)\right) \rmd^2 z =
\tau\left(\lambda_j(\btheta)\right),
\]
where $\rmd^2 z= \rmd u \, \rmd v$ and $\tau(z)$ is an infinitely
differentiable complex function whose partial derivatives with
respect to $u$ and $v$ decrease faster than any power of
$1/\abs{z}$.  Moreover,
\[
\bdelta\left(z - \lambda_j(\btheta)\right)=\delta\left(u - \rpart
\left(\lambda_j(\btheta)\right)\right)\delta\left(v - \ipart
\left(\lambda_j(\btheta)\right)\right)
\]
is the product of two Dirac delta functions with real arguments.
Then, we have
\[
\fl \int_{\mathbb{C}} \tau(z) \rho(z) \, \rmd^2 z =
\frac{\Omega(N)}{N-1}\sum_{j=1}^{N -1}\int_{[0,\,2\pi]^N}
\int_{\mathbb{C}}\tau(z)\, \bdelta\left(z -
\lambda_j(\btheta)\right) \Delta^2(\btheta) \rmd^2 z \, \rmd^N
\btheta,
\]
 where the distribution
\begin{eqnarray}
\label{dens} \rho(z):= \frac{\Omega(N)}{N-1}
\sum_{j=1}^{N-1}\int_{[0,\,2\pi]^N}\bdelta\left(z -
\lambda_j(\btheta)\right)\Delta^2(\btheta) \rmd^N \btheta
\end{eqnarray}
defines the density of $\{\lambda_{j}(\btheta)\}_{j=1}^{N-1}$ .

The main tool that we shall use to evaluate~\eref{dens} is a basic
identity that expresses {\it Toeplitz determinants} in terms of
integrals over the unitary group.  If
\[
f(\theta)=\sum_{k=-\infty}^{\infty}\hat{f}_k \, \rme^{\rmi k \,
\theta}
\]
is a complex function on the unit circle, then we denote by
$D_{N-1}[f]$ the determinant of the {\it Toeplitz matrix}
\[
T_{N-1}[f] := \left( \begin{array}{cccc}
\hat{f}_0 & \hat{f}_1 & \cdots & \hat{f}_{N-1} \\
\hat{f}_{-1} & \hat{f}_0 & \cdots & \hat{f}_{N-2} \\
\vdots & \vdots & &  \vdots \\
\hat{f}_{-(N-1)} & \hat{f}_{-(N-2)} & \cdots & \hat{f}_0
\end{array} \right).
\]
Now, let $\Phi_{f}$ be a {\it class function}, i.e. a complex
function on $\UN$ such that
\[
\Phi_{f}\left(VUV^{-1}\right) = \Phi_{f}(U), \quad V, U \in \UN.
\]
Furthermore, suppose that
\begin{equation}
\label{pr_f} \Phi_{f}(U)= f(\theta_1)f(\theta_2) \cdots
f(\theta_N),
\end{equation}
where $\{\rme^{\rmi \theta_j} \}_{j=1}^N$ are the eigenvalues of
$U$. The {\it Heine-Szeg\H{o} identity}~\cite{Sze67} states that
\begin{eqnarray}
\label{H-S_id}
D_{N-1}[f]&= \int_{\UN} \Phi_f(U)\, \rmd \mu (U) \nonumber \\
&= \Omega(N)\int_{[0,\,2\pi]^N} \left(\prod_{j=1}^N
f(\theta_j)\right)\Delta^2(\btheta) \rmd^N \btheta.
\end{eqnarray}

In order to apply this formula, we must express the sum of delta
functions in~\eref{dens} as a product of the form~\eref{pr_f}. The
zeros of $Z\,'(U,z)$ that are not multiple roots of $Z(U,z)$ are
the same as those of the logarithmic derivative of $Z(U,z)$. Since
the set of unitary matrices with degenerate eigenvalues has zero
measure, we rewrite the integrand in equation~\eref{dens} as
\begin{equation}
\label{ndel}
 \sum_{j=1}^{N-1}
\bdelta\left(z-\lambda_j\right)=\bdelta
\left(Z\,'(U,z)/Z(U,z)\right)\abs{\frac{\rmd}{\rmd z}
\left[Z\,'(U,z)/Z(U,z)\right]}^2.
\end{equation}
In the next step we use the integral representation of a delta
function:
\[
\delta(x) = \frac{1}{2\pi}\int_{-\infty}^{\infty}\rme^{\rmi \xi x}
\, \rmd \xi.
\]
The complex delta function in the right-hand side of
equation~\eref{ndel} now becomes
\begin{equation}
\label{intdelt}
 \fl \bdelta
\left(Z\,'(U,z)/Z(U,z)\right)=\frac{1}{4\pi^2} \int_{\mathbb{C}}
\exp \left[\frac{\rmi}{2}\left(\frac{Z\,'(U,z)}{Z(U,z)} \;
\overline w + \frac{\overline{Z\,'(U,z)}}{\overline{Z(U,z)}}\, w
\right)\right] \rmd^2 w.
\end{equation}
Clearly, the identity~\eref{H-S_id} can be applied to the argument
of the integral~\eref{intdelt}; furthermore, the Jacobian in
equation~\eref{ndel} can be transformed into a product of the
form~\eref{pr_f} by using the following representation of the
modulus square of a complex number:
\[
\abs{z}^2 = - \left.\frac{\partial^2}{\partial \alpha^2} \;
G(\alpha,z)\right |_{\alpha=0}, \quad \alpha \in \mathbb{R},
\]
where
\[
G(z,\alpha):=\exp\left[\rmi \alpha \left(z +
\overline{z}\,\right)/2\right]+ \exp\left[\alpha \left(z -
\overline{z}\,\right)/2\right].
\]
Finally, the density~\eref{dens} becomes
\begin{eqnarray}
\label{ndens} \fl  \rho(z) = - \frac{1}{4\pi^2\left(N-1\right)}
\nonumber \\
\fl \qquad \quad \times \frac{\partial^2}{\partial \alpha^2} \left
[\int_{\mathbb{C}}\Bigl( D_{N-1}[\exp(\rmi g)](w,\alpha,z) +
D_{N-1}[\exp(\rmi h)](w,\alpha,z) \Bigr)\rmd^2 w
\right]_{\alpha=0},
\end{eqnarray}
where
\begin{equation*}
\label{g&h}
  g(\theta;w,\alpha,z):= \frac{1}{2}
\left(\frac{\overline w}{z - \rme^{\rmi \theta}} +
\frac{w}{\overline z - \rme^{-\rmi \theta}}\right) -
\frac{\alpha}{2} \left(\frac{1}{\left(z - \rme^{\rmi
\theta}\right)^2} + \frac{1}{\left(\overline z - \rme^{-\rmi
\theta}\right)^2} \right)
\end{equation*}
and
\begin{equation*}
  h(\theta;w,\alpha,z) := \frac{1}{2}
\left(\frac{\overline w}{z - \rme^{\rmi \theta}} +
\frac{w}{\overline z - \rme^{-\rmi \theta}}\right) + \frac{\rmi
\alpha}{2} \left(\frac{1}{\left(z - \rme^{\rmi \theta}\right)^2} -
\frac{1}{\left(\overline z - \rme^{-\rmi \theta}\right)^2} \right)
\end{equation*}
are real functions.

\section{Asymptotics of $\rho(z)$}
\label{asrho}

Computing~\eref{ndens} exactly appears to be a formidable task.
However, there exists a powerful theorem of Szeg\H{o} that will
allow us to compute the leading-order asymptotics of $\rho(z)$ as
$N \rightarrow \infty$.

\vspace{5pt}
 \noindent {\bf The strong Szeg\H{o} limit theorem.}
{\it Let}
\begin{equation*}
\label{sigma} \eta(\theta)= \sum_{k=-\infty}^{\infty} \hat{\eta}_k
\, \rme^{\rmi k \, \theta}
\end{equation*}
{\it be a complex function on the unit circle. If the series}
\begin{equation}
\label{conditions}
 \sum_{k=-\infty}^{\infty} \abs{\hat{\eta}_k}
\quad {\it and} \quad \sum_{k=-\infty}^{\infty} \abs{k}
\abs{\hat{\eta}_k}^{\, 2}
\end{equation}
{\it converge, then}
\begin{equation}
\label{Sf} D_{N-1}[\exp(\eta) ] = \exp\left(\hat{\eta}_0 N +
\sum_{k=1}^{\infty}k\, \hat{\eta}_{-k}\hat{\eta}_k + {\rm
o}(1)\right), \quad N \rightarrow \infty.
\end{equation}
The first proof of this theorem was given by Szeg\H{o}
in~\cite{Sze52} under stronger conditions; several proofs have
since been developed~\cite{Bax63,BD02}.

In the region $\abs{z}< 1$, $g(\theta;w,\alpha,z)$ and
$h(\theta;w,\alpha,z)$ are just sums of geometric series and of
their derivatives whose Fourier coefficients can be easily
computed. We have
\[
g(\theta; w,\alpha,z) = \sum_{k=-\infty}^{\infty} \hat{g}_k \,
\rme^{\rmi k \theta} \quad {\rm and} \quad h(\theta; w,\alpha,z) =
\sum_{k=-\infty}^{\infty} \hat{h}_k \, \rme^{\rmi k \theta},
\]
where
\begin{equation*}
\label{gfc}
\hat{g}_0=0, \quad \hat{g}_k = -\frac{w}{2}
\overline{z}^{\;k-1} - \frac{\alpha}{2} \left(k-1\right)
\overline{z}^{\;k-2}, \quad k \in \mathbb{Z}^+
\end{equation*}
and
\begin{equation*}
\label{hfc}
 \hat{h}_0=0, \quad \hat{h}_k =
-\frac{w}{2}\overline{z}^{\;k-1} - \frac{\rmi
\alpha}{2}\left(k-1\right)\overline{z}^{\;k-2}, \quad k \in
\mathbb{Z}^+.
\end{equation*}
Since $g(\theta;w,\alpha,z)$ and $h(\theta;w,\alpha,z)$ are real,
$\hat{g}_{-k}=\overline{\hat{g}}_k$ and $\hat{h}_{-k}=
\overline{\hat{h}}_k$. Computing the argument in the exponential
of equation~\eref{Sf} involves only summing and differentiating
geometric series.  We obtain
\numparts
\begin{eqnarray}
E(g)=\sum_{k=1}^\infty k\, \abs{\hat{g}_{k}}^2 & = c(w,\alpha,r) +
\frac{\alpha \rpart (w\overline{z})}{\left(1 - r^2\right)^3}, \\
E(h) = \sum_{k=1}^\infty k\, \abs{\hat{h}_{k}}^2 & = c(w,\alpha,r)
+ \frac{\alpha \ipart (w\overline{z})}{\left(1 - r^2\right)^3},
\end{eqnarray}
\endnumparts
where $r=\abs{z}$ and
\[
c(w,\alpha,r)=\frac{\abs{w}^{\,2}}{4}
\frac{1}{\left(1-r^2\right)^2} +\frac{\alpha^2}{2}
\left[\frac{3r^2}{\left(1-r^2\right)^4} +
\frac{1}{\left(1-r^2\right)^3}\right].
\]
As a consequence, the second sum in~\eref{conditions} is finite.
Furthermore, we have
\[
\lim_{k \rightarrow \infty}
\abs{\frac{\hat{g}_{k+1}}{\hat{g}_k}}=\lim_{k\rightarrow \infty}
\abs{\frac{\hat{h}_{k+1}}{\hat{h}_k}} = r < 1.
\]
Hence, the first series in~\eref{conditions} converges too, and
the strong Szeg\H{o} limit theorem applies.

When computing derivatives of the asymptotics of the
integral~\eref{ndens}, care must be taken that the error term does
not become comparable or even greater than the leading-order term.
This could happen, for example, if the remainder were a highly
oscillatory function of $\alpha$. It turns out that the
convergence of $D_{N-1}[\exp(\rmi g)]$ and $D_{N-1}[\exp(\rmi h)]$
to~\eref{Sf} is so fast that the derivatives of the error term
remain small. This is proved in the appendix.

\begin{figure}
\centering \subfigure[]{ \label{fig3a}
\includegraphics[width=3.0in]{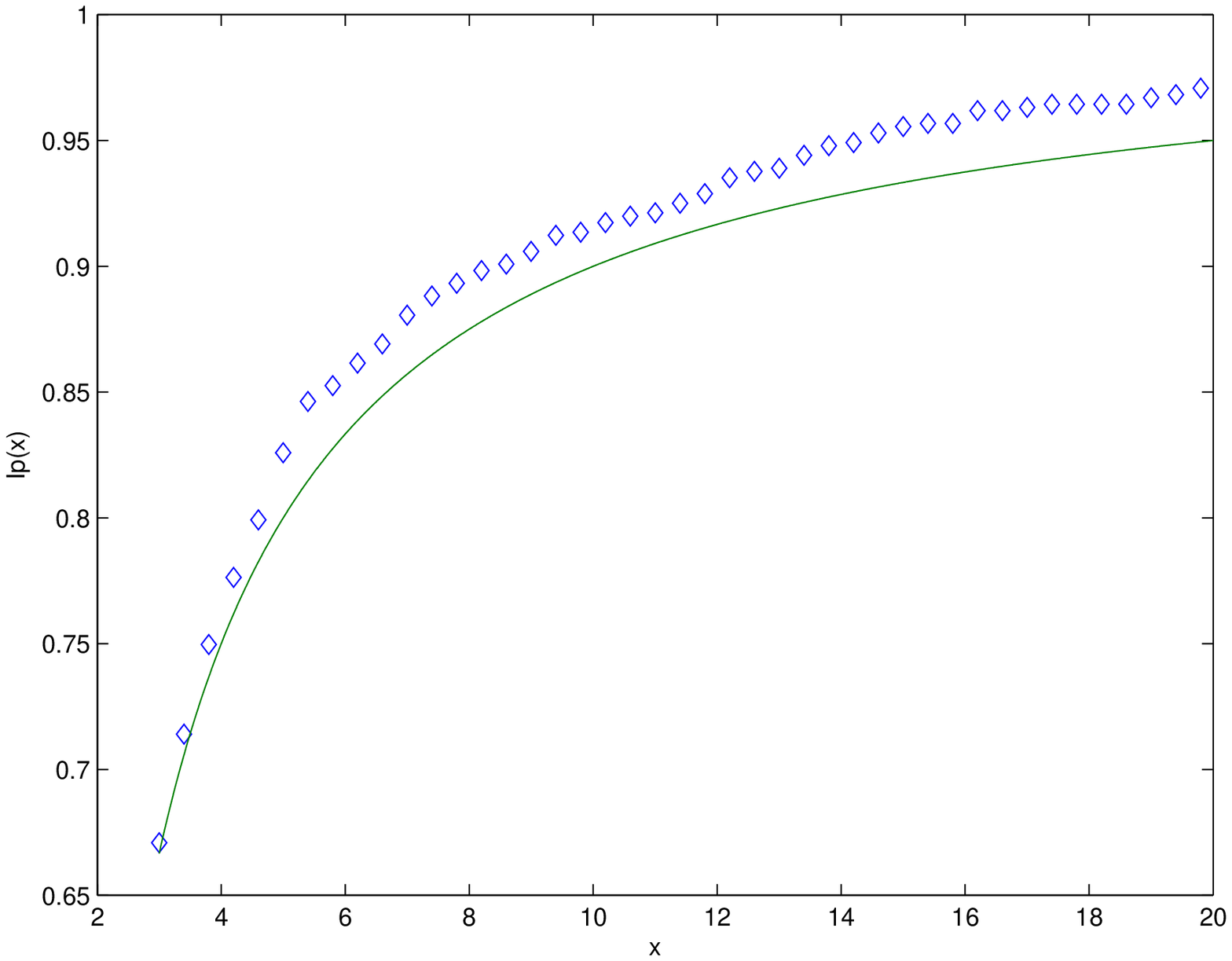}}
 \subfigure[]{\label{fig3b}
  \includegraphics[width=3.0in]{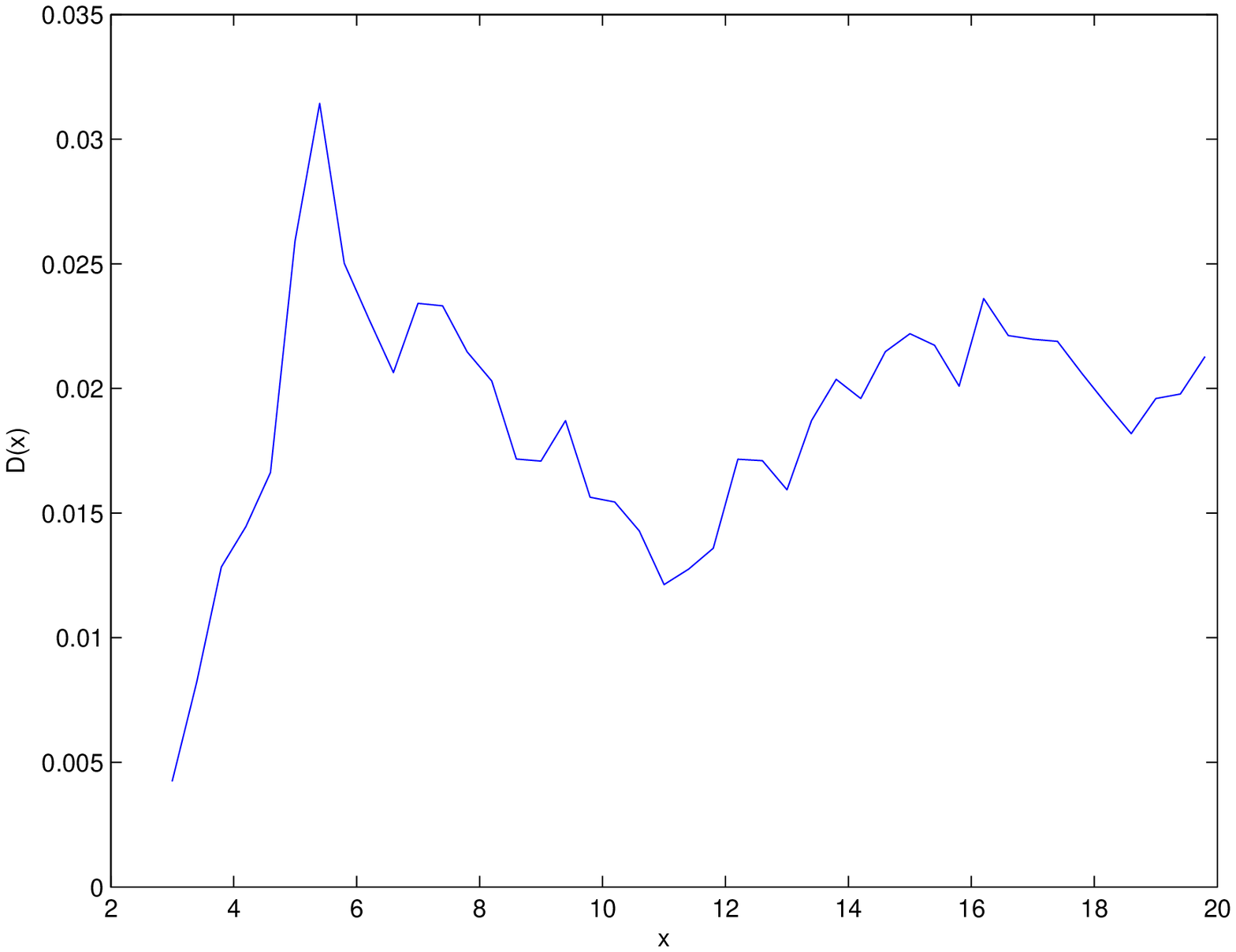}}
\caption{Comparison (a) and difference (b) between $\Ip(x)$
computed numerically for $N=800$ (\opendiamond) and
formula~\eref{mr} (\full).}
      \label{fig3}
\end{figure}
In equation~\eref{ndens} the second derivative commutes with the
integral, hence differentiating twice with respect to $\alpha$
gives
\begin{eqnarray*}
\fl \frac{\partial^2}{\partial \alpha^2} \Biggl [D_{N-1}[\exp(\rmi
g)](w,\alpha,z) + D_{N-1}[\exp(\rmi h)](w,\alpha,z)
 \Biggr]_{\alpha=0} \\
\sim  \left[\frac{\abs{w}^2 r^2}{\left(1-r^2\right)^6} -
 \frac{6r^2}{\left(1-r^2\right)^4}
 -\frac{2}{\left(1-r^2\right)^3}\right]\exp\left[-
 \frac{\abs{w}^2}{4\left(1-r^2\right)^2}\right], \quad N \rightarrow
 \infty.
\end{eqnarray*}
 This expression can be trivially integrated.  Finally, we obtain
\begin{equation*}
\label{fdens} \rho(z) \sim
\frac{2}{\pi\left(N-1\right)\left(1-r^2\right)^2}\, , \quad
N\rightarrow \infty.
\end{equation*}
As anticipated, $\rho(z)$ depends only on the distance from the
origin $r$ and is asymptotically concentrated  in a small region
near the unit circle, which explains the migration of the roots of
$Z\,'(U,\lambda)$ observed in figure~\ref{fig1} as $N$ increases.
Since $\rho(z)$ is a density, it must be normalized to one,
therefore we require
\[
\int_0^{2\pi} \int_0^{1 - \epsilon} r \rho(z)\, \rmd r \rmd \phi
\sim 1, \quad N \rightarrow \infty, \quad \epsilon \rightarrow 0.
\]

Let us now define $\Ip(x)$ to be the fraction of the zeros in the
annulus \newline $1 - x/(N-1) \le r < 1$, where $x={\rm o}(N)$,
i.e.
\begin{eqnarray}
\label{mr}
\fl \Ip(x)  = 1- \int_{0}^{2\pi} \int_0^{1-x/(N-1)}
r\rho(z)\, \rmd r \rmd \phi \sim 1 -  \frac{2}{N-1}
\left[\frac{1}{1-r^2}\right]_0^{1-x/(N-1)} \nonumber \\
\lo\sim 1-1/x, \quad N\rightarrow \infty, \quad x \rightarrow
\infty.
\end{eqnarray}
 As $N \rightarrow \infty$ the leading-order term of $\Ip(x)$ is
independent of $N$. \Eref{mr} is the main result of this section.
In figure~\ref{fig3} formula~\eref{mr} is compared with $\Ip(x)$
computed for a matrix of dimension $N=800$.

\section{The asymptotics $x \rightarrow 0$}
\label{smxas} The small $x$ asymptotics of $\Ip(x)$ requires first
the evaluation of the limit $x \rightarrow 0$ and then of the
limit $N \rightarrow \infty$.  Szeg\H{o}'s theorem gives an
asymptotic expression as $N \rightarrow \infty$ before the limit
$x \rightarrow 0$ can be taken, therefore provides information
only for relatively large $x$.  In order to determine $\Ip(x)$ in
the limit $x\rightarrow 0$, the integral in equation~\eref{ndens}
needs to be evaluated for finite $N$. Such computation seems to be
an extremely difficult task: the integrand has essential
singularities that usual techniques in RMT and complex analysis
cannot tackle. Notwithstanding such obstacles, it turns out that
$\Ip(x)$ can be derived in the limit $x \rightarrow 0$ with the
help of a heuristic argument.

As was mentioned in section~\ref{zUp}, from figure~\ref{fig1} it
appears that if $\rme^{\rmi \theta_j}$ and $\rme^{\rmi
\theta_{j+1}}$ are two consecutive roots of $Z(U,z)$ which are
close to each other, then as $N \rightarrow \infty$ there is often
a root of $Z\,'(U,z)$ near the midpoint~\eref{mpoin}. Thus, one
might be led to assume that for small $x$ the distance from the
origin $r$ is distributed like
\begin{eqnarray}
\label{radius}
1 - x/(N- 1) & =\abs{\frac{\rme^{\rmi \theta_j} +
\rme^{\rmi \theta_{j +1}}}{2}} =
 \frac{\sqrt{2 + 2 \cos \left(2 \pi
S/N\right)}}{2} \nonumber \\
 & \approx 1 - \frac{\pi^2 S^{\,2}}{2N^2}, \quad 1\le j \le N,
\end{eqnarray}
where $S$ is the rescaled distance, or spacing, between phases of
consecutive eigenvalues, i.e.
\[
S= \frac{N}{2\pi}\abs{\theta_{j+1} - \theta_j}, \quad 1 \le j \le
N.
\]
This is trivially true only for $N=2$;  since $S=\Or(1)$, for $N >
2$ equation~\eref{radius} would imply an average distance of the
zeros of $Z\,'(U,z)$ from the unit circle of order $1/N^2$, and
therefore a dependence of $\Ip(x)$ on $N$ even at the leading
order, which contradicts the numerics reported in
figure~\ref{fig2} and formula~\eref{mr}.

\begin{figure}
\centering
\includegraphics[width=3.0in]{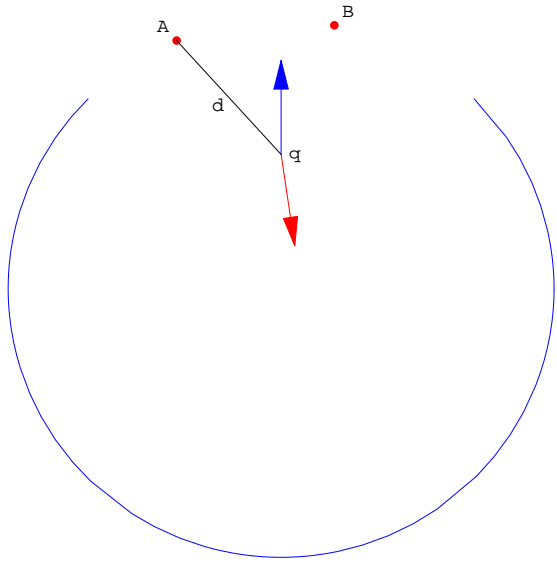}
\caption{Schematic representation of the two components
$\mathbf{E}_\mu(q)$ and $\mathbf{E}_{{\rm AB}}(q)$ of the electric
field~\eref{efd} at a point $q$ close to the unit circle in
Dyson's electrostatic model.}
\label{elecm}
\end{figure}

It turns out that behaviour of $\Ip(x)$ as $x \rightarrow 0$ can
be understood using Dyson's electrostatic model for the CUE
ensemble (see, e.g.,~\cite{Meh91}). The zeros of $Z(U,z)$ can be
interpreted as $N$ unit charges confined in a two-dimensional
universe and moving in a thin circular conducting wire of radius
one. The electric field generated at any point in the complex
plane by this Coulomb gas is just the complex conjugate of the
logarithmic derivative of $Z(U,z)$, i.e.
\[
\mathbf{E}(z)= \frac{1}{\overline{z} - \rme^{-\rmi \theta_1}} +
\frac{1}{\overline{z} - \rme^{-\rmi \theta_2}} + \cdots +
\frac{1}{\overline{z}- \rme^{-\rmi \theta_N}}.
\]
Hence, the zeros of $Z\,'(U,z)$ are located where $\mathbf{E}(z)$
vanishes.  Now, the field at point $q$, whose distance from the
unit circle is of order $1/N$, can be separated into two
components: the first one is the field of the two charges closest
to $q$, let us denote them by A and B, whose strength is clearly
of order $N$; the second one is the field generated by the other
$N - 2$ charges. As $N$ increases, $q$ approaches the unit circle,
and the latter component of $\mathbf{E}(q)$ can be approximated by
the field of a continuous circular charge distribution with
density $\mu=N/(2\pi)$, i.e.
\begin{equation}
\label{efd} \mathbf{E}(q) \approx \mathbf{E}_\mu(q) +
\mathbf{E}_{{\rm AB}}(q),
\end{equation}
where $\mathbf{E}_\mu(q)$ is the field generated by $\mu$ and
$\mathbf{E}_{{\rm AB}}(q)$ the one determined by A and B. For
large $N$,  $\mathbf{E}_\mu(q)$ and $\mathbf{E}_{{\rm AB}}(q)$
have approximately opposite directions, thus $\mathbf{E}(z)$ might
vanish at $q$ only if $\abs{\mathbf{E}_\mu(q)}=\Or(N)$. This
situation is described schematically in figure~\ref{elecm}.
Determining the order of magnitude of $\mathbf{E}_\mu(q)$ is a
simple exercise in electrostatics. If the continuous charge
distribution filled the whole unit circle, the field inside it
would be zero.  Thus, by linearity $\mathbf{E}_\mu(q)$ is equal
and opposite to the field of a circular arc $[-\tilde \theta,
\tilde \theta]$ with charge density $\mu$ and containing no more
than four eigenvalues. We have
\begin{equation*}
\label{el_fil}
\abs{\mathbf{E}_\mu(q)} = \frac{N}{2\pi}
\int_{-\tilde\theta}^{\tilde\theta} \, \frac{1}{d(\theta)}\sqrt{1-
\frac{\sin^2 \theta}{d(\theta)^2}} \, \rmd \theta,
\end{equation*}
where $d(\theta)$ is the distance between $q$ and $\rme^{\rmi
\theta}$. By setting $d(\theta)=t(\theta)/N$, with
$t(\theta)=\Or(1)$, and applying the mean value theorem we obtain
\[
\abs{\mathbf{E}_\mu(q)}= \frac{\tilde \theta N^2}{\pi
t(\xi)}\sqrt{1-\frac{\sin^2\xi}{d(\xi)^2}}, \quad -\tilde \theta
\le \xi \le \tilde \theta.
\]
Since the length of the interval $[-\tilde \theta, \tilde \theta]$
is of the order of few level spacings,  $\tilde \theta = \Or(1/N)$
and $\abs{\mathbf{E}_\mu(q)}=\Or(N)$.

The field $\mathbf{E}_{{\rm AB}}(z)$ vanishes at the
midpoint~\eref{mpoin}, but for large $N$ the presence of
$\mathbf{E}_\mu(z)$ shifts the zeros of $\mathbf{E}(z)$ at a
distance of order $1/N$ from the unit circle. Since at first
approximation the contribution of $\mathbf{E}_\mu(z)$ does not
depend on the spacing $S$, from~\eref{radius} it is reasonable to
assume that as $x \rightarrow 0$ the distance of the zeros of
$Z\,'(U,z)$ from the unit circle is approximately distributed like
$S^{\,2}/N$. Hence, we shall conjecture that
\begin{equation}
\label{conj2}
x \sim \beta\, \frac{\pi^2 S^{\,2}}{2}\, , \quad x
\rightarrow 0,
\end{equation}
where $\beta$ is a constant independent of $N$.

From equation~\eref{conj2} it is straightforward to derive an
asymptotics expression for $\Ip(x)$ as $x \rightarrow 0$. We have
\[
 \Ip(x) \sim \int_0^x \gamma(y)\, \rmd y, \quad x \rightarrow 0,
\]
where
\[
 \gamma(y):= \frac{1}{\pi}\sqrt{\frac{1}{2\beta y }}\, P_{{\rm
 CUE}}\left[\frac{1}{\pi}
 \left(\frac{2y}{\beta}\right)^{1/2} \right]
\]
and $P_{{\rm CUE}}(S)$ is the CUE spacing distribution in the
limit $N \rightarrow \infty$, i.e.
\[
P_{{\rm CUE}}(S):= \lim_{N\rightarrow \infty} P_{{\rm CUE}}(N;S).
\]
$P_{{\rm CUE}}(S)$ has a power series expansion with infinite
radius of convergence:
\begin{equation}
\label{pcue}
 P_{{\rm CUE}}(S)=  \sum_{l=0}^{\infty}
(l+4)(l+3)E_l\, S^{2 + l}.
\end{equation}
\begin{figure}
\centering \subfigure[]{ \label{fig4a}
\includegraphics[width=3.0in]{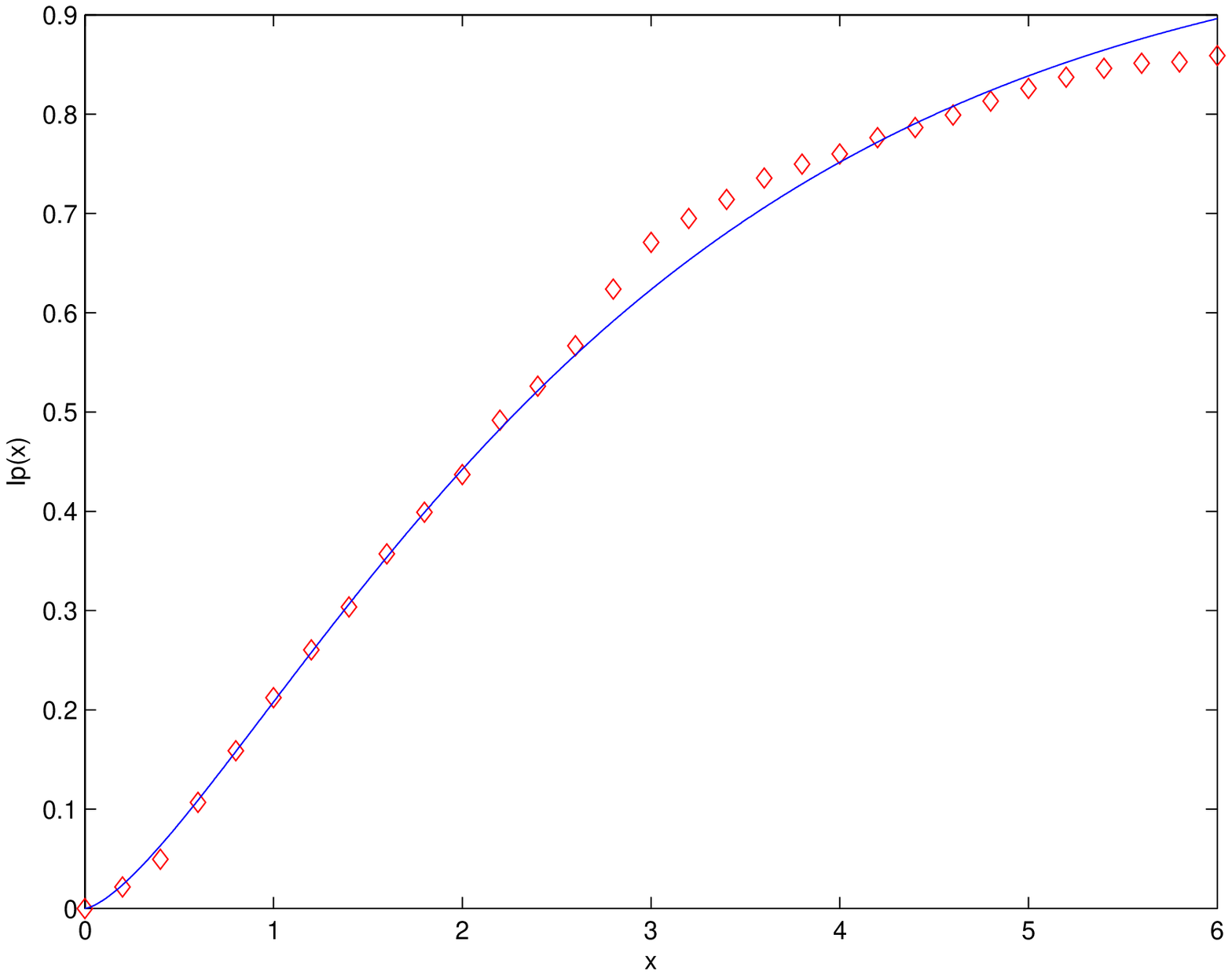}}
\subfigure[]{\label{fig4b}
  \includegraphics[width=3.0in]{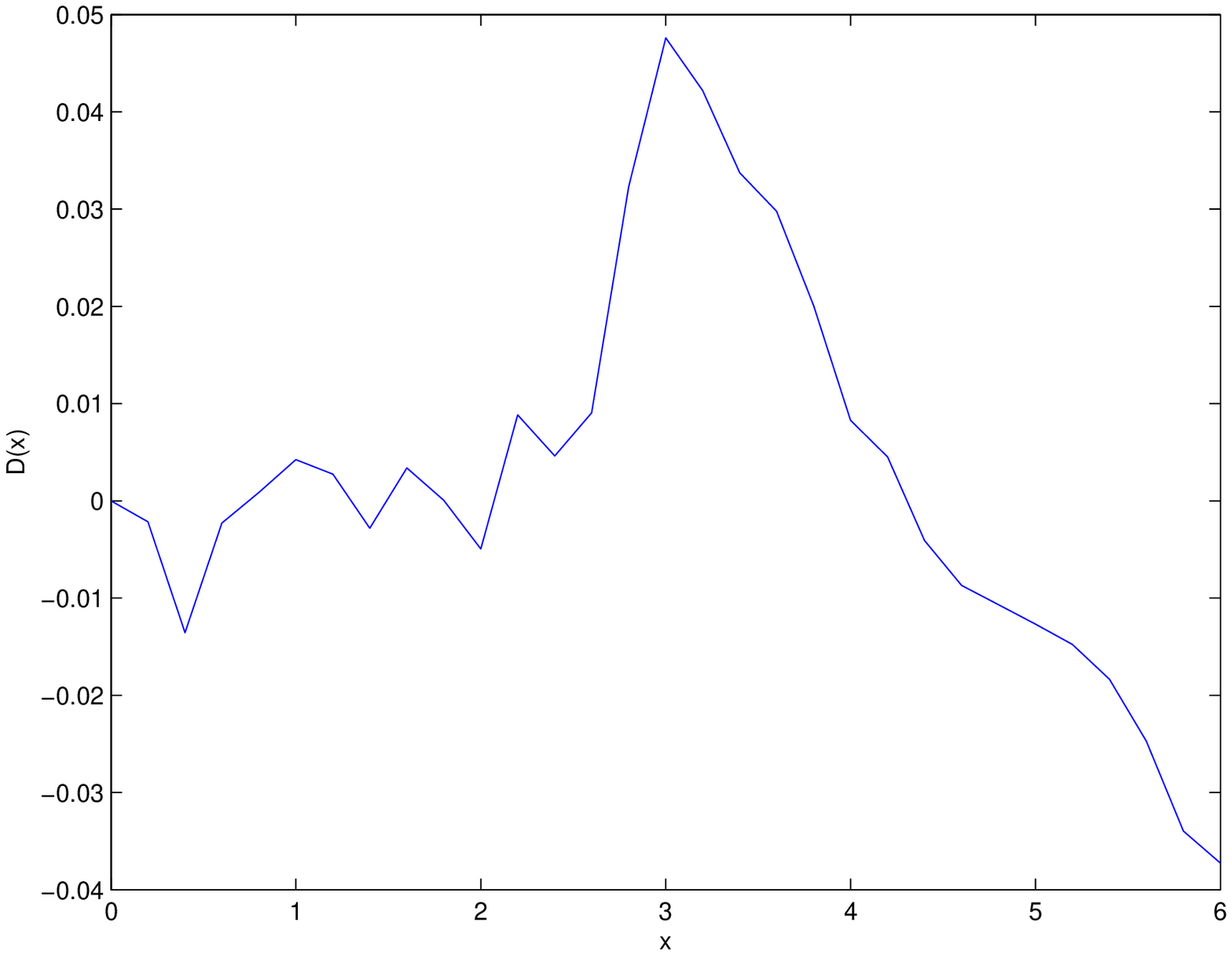}}
\caption{Comparison (a) and difference (b) between $\Ip(x)$
computed numerically for $N=800$ (\opendiamond) and the one
defined by the series~\eref{conj} truncated at $l=30$ and with
$\beta=1/2$ (\full).} \label{fig4}
\end{figure}
There exist efficient algorithms for computing the coefficients
$E_l$ (see, e.g.,~\cite{Hak00}), which with symbolic mathematical
packages can be evaluated exactly up to very high values of $l$.
Using~\eref{pcue} one can easily obtain a series expansion for
$\Ip(x)$:
\begin{equation}
\label{conj}
 \Ip(x) \sim \sum_{l=0}^{\infty}\left(\frac{2}%
 {\beta \pi^2}\right)^{\frac{l + 3}{2}}(l+4)\; E_l\;
x^{\frac{l+3}{2}}, \quad x \rightarrow 0.
\end{equation}
Now we have to determine the parameter $\beta$, which can be found
only empirically. It turns out that if we set $\beta=1/2$ there
exists an astonishing agreement between~\eref{conj} and numerics
in the region where the large $x$ asymptotics is not valid. This
is shown in figure~\ref{fig4}.

Notwithstanding the accuracy with which the series~\eref{conj}
models the numerical data, it can be expected to approximate
$\Ip(x)$ only for small $x$; indeed, it tends to one much faster
than $1 - 1/x$. Furthermore, in deriving~\eref{conj} we have
implicitly assumed that the main contribution to $\Ip(x)$ comes
only from the two zeros of $Z(U,z)$ closest to a given root of
$Z\,'(U,z)$; we now need to estimate in what region this
assumption is justified.

Let us consider $k + 2$ successive eigenvalues of a unitary
matrix. Since Haar measure is translation invariant on the unit
circle, the corrections to~\eref{conj} will depend by all possible
rescaled distances
\[
S_k = \frac{N}{2\pi}\abs{\theta_{j + k + 1} - \theta_j}, \quad 1
\le j \le N.
\]
However, as $S_k \rightarrow 0$, the limit densities $P_{{\rm
CUE}}(S_k)$ go to zero very fast, hence the dependence on $S_k$
does not affect the first few terms of the series~\eref{conj}. For
example, consider three consecutive zeros of $Z(U,z)$ close to
each other, say
\begin{equation*}
\fl \exp\left(\frac{2\pi \rmi}{N}\, \alpha \right), \quad
\exp\left(\frac{2\pi \rmi}{N}\, \left( \alpha + S_0 \right)\right)
\quad {\rm and}\quad  \exp\left(\frac{2\pi \rmi}{N} \,\left(
\alpha + S_1 \right)\right), \quad S_1 \ge S_0 \ge 0.
\end{equation*}
Simple algebra and Dyson's model indicate that the rescaled
distance $x$ should be distributed like
\[
\beta' (S_0 + S_1)^2,
\]
for some constant $\beta'$ independent of $N$. It turns out  that
\begin{equation}
\label{ds1}
 P_{{\rm CUE}}(S_1)  = \frac{\pi^6 S_1^{\,7}}{4050}
+\Or(S_1^{\,8}),
\end{equation}
which suggests that the  contributions to~\eref{conj} due to $S_1$
leave the coefficients of the terms $x^d$ with $d < 4$ unchanged.
For $k>1$, $P_{{\rm CUE}}(S_k)$ goes to zero as $S_k \rightarrow
0$ even faster than~\eref{ds1}. These considerations and
formula~\eref{conj} give the following expression for the
asymptotic expansion of $\Ip(x)$:
\[
 \Ip(x) = \frac{8}{9\pi}\; x^{3/2} - \frac{64}{225\pi}\;x^{5/2} +
\frac{128}{2205\pi}\;x^{7/2} + \Or(x^4).
\]
\begin{figure}
\centering \subfigure[]{\label{mnum}
\includegraphics[width=3.0in]{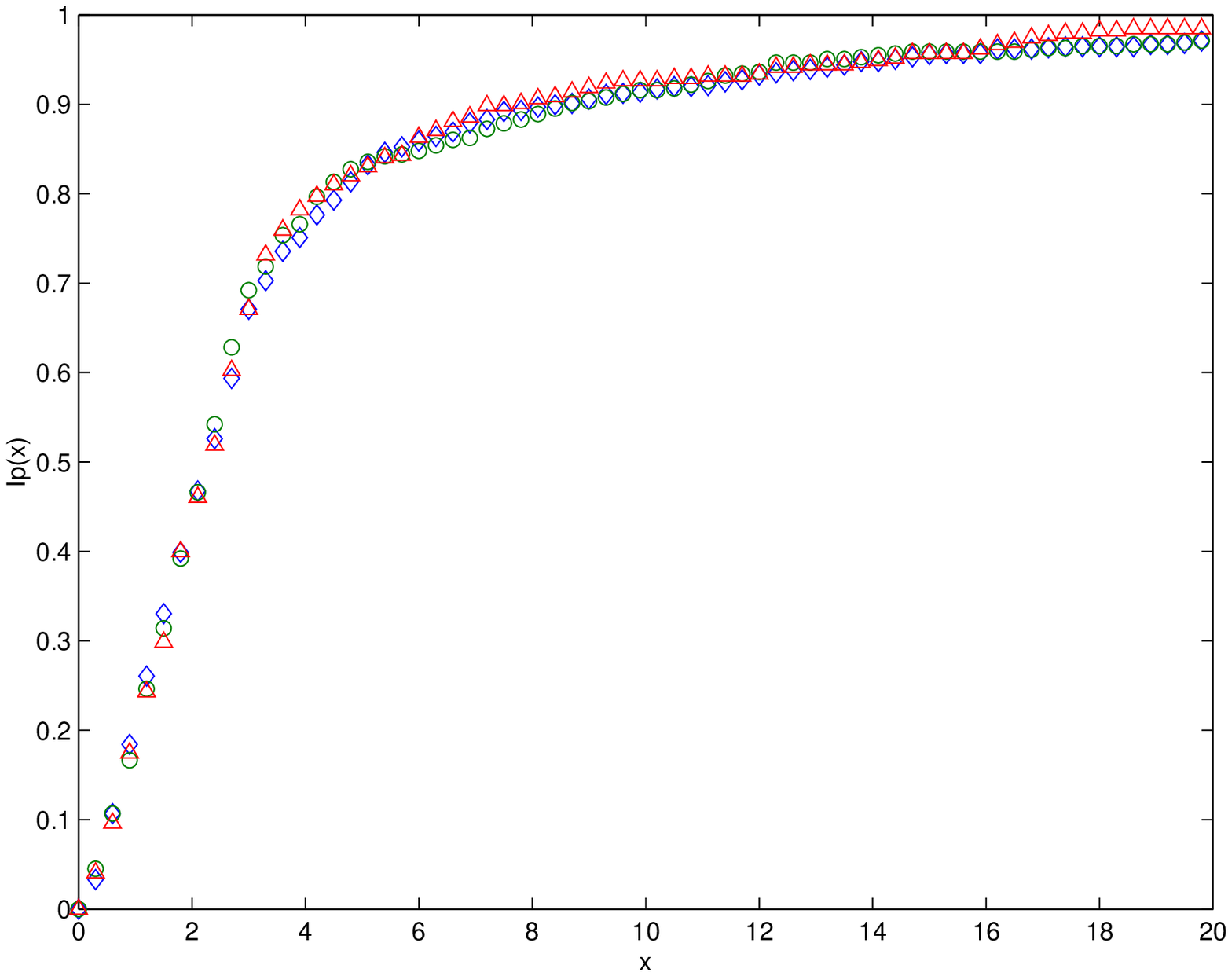}}
\subfigure[]{\label{bcurves}
\includegraphics[width=3.0in]{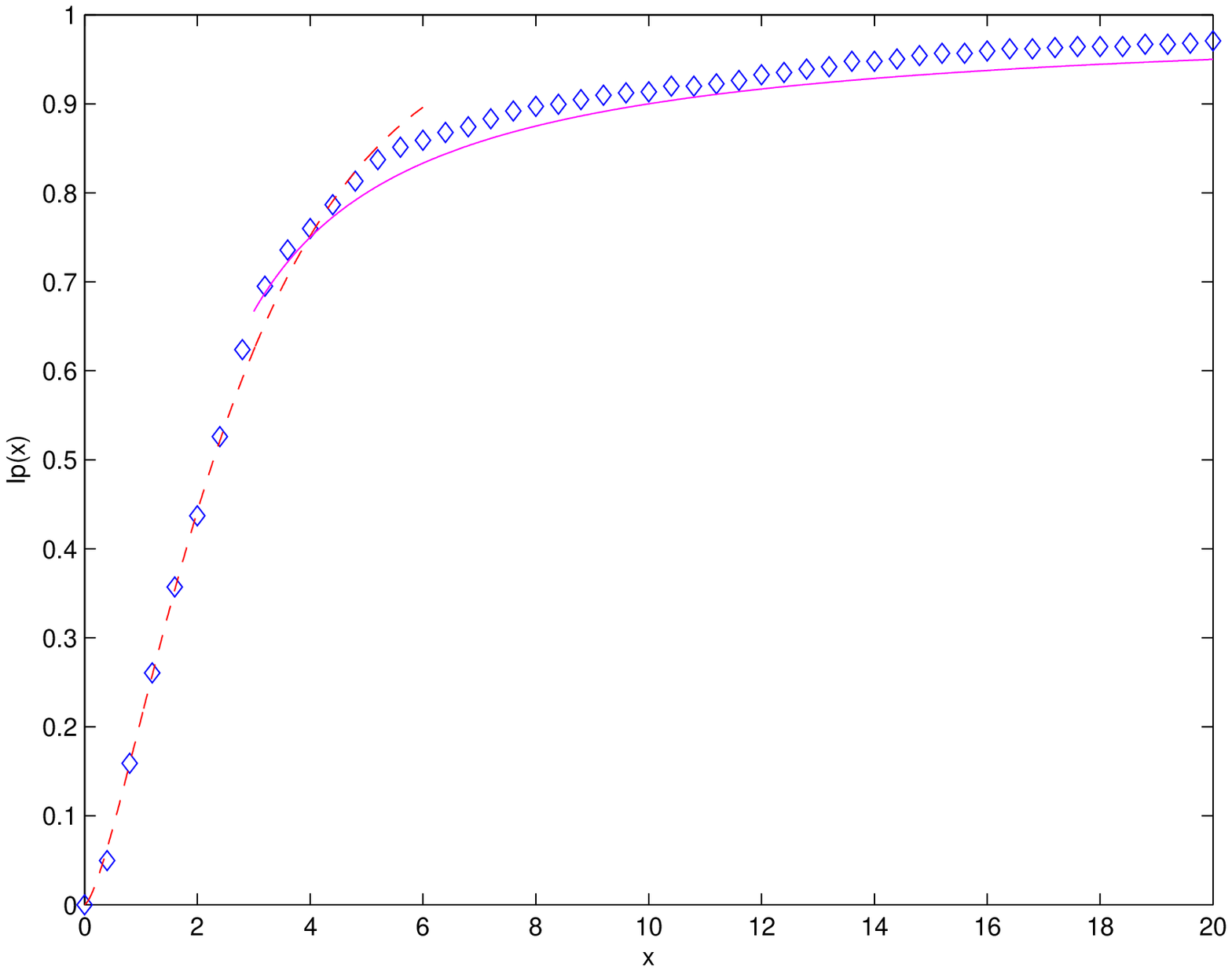}}
\caption{(a) Fraction of zeros of $Z\,'(U,z)$ in the region
$1-x/(N-1) \le \abs{z} < 1$ for $N=800$ (\opendiamond), $N=500$
(\opencircle) and $N=400$ (\opentriangle). (b) Same data as in (a)
for $N=800$ (\opendiamond) compared with formula~\eref{mr} (\full)
and with the series~\eref{conj} truncated at $l=30$ and with
$\beta=1/2$ (\dashed).}
      \label{fig2}
\end{figure}
However, from the agreement with numerics observed in
figure~\ref{fig4} and~\ref{bcurves}, we would expect that the
corrections to the series~\eref{conj} should be negligible up to
terms of order much higher than $x^4$.

The integrated distribution $\Ip(x)$ is plotted in
figure~\ref{mnum} for zeros of $Z\,'(U,z)$ computed numerically
for random unitary matrices of various dimensions; clearly,
$\Ip(x)$ tends to a limit function. Figure~\ref{bcurves} shows
that equation~\eref{mr} and the series~\eref{conj} together,
although asymptotic formulae, approximate $\Ip(x)$ with high
accuracy for all $x\ge 0$.

\section{Concluding remarks}
\label{concl}

We have investigated the density $\rho(z)$ of the roots of
$Z\,'(U,z)$, where $Z(U,z)$ is the characteristic polynomial of a
random $N \times N$ unitary matrix with distribution given by Haar
measure on the unitary group.  Since the locally determined
statistical properties of the Riemann zeta function high up the
critical line can be modelled by $Z(U,z)$, it is expected that
$\rho(z)$ will accurately describe the behaviour of the zeros of
$\zeta\,'(s)$. It turns out that instead of $\rho(z)$, it is more
convenient to study $\Ip(x)$, the fraction of the roots in the
region $1 - x/(N-1)\le \abs{z} < 1$. In the analogous problem for
the zeta function, this is equivalent to looking at the fraction
of the zeros of $\zeta\,'(s)$ in the region $1/2 < \sigma \le 1/2
+ x/\log T$ where $\sigma = \rpart(s)$ and $T$ is the height on
the critical line. It is shown that as $N\rightarrow \infty$,
$\Ip(x)$ becomes independent of $N$.

The density $\rho(z)$ can be defined as the average over $\UN$ of
the sum of delta functions
\[
\frac{1}{N-1}\sum_{j=1}^{N-1}
\bdelta\left(z-\lambda_j(\btheta)\right),
\]
where the $\lambda_j(\btheta)$ are the zeros of $Z\,'(U,z)$. The
behaviour of $\Ip(x)$ for large $x$ can be computed by applying
standard techniques for integrals over $\UN$. The sum of Dirac
deltas can be manipulated in such a way that eventually the
average over $\UN$ is reduced to the computation of the second
derivative of an integral over the complex plane of the sum of two
Toeplitz determinants. Furthermore, the integrand satisfies the
hypothesis of the strong Szeg\H{o} limit theorem, which gives a
simple asympotic expression for such determinants. Further simple
manipulations lead to
\[
\Ip(x) \sim 1 - 1/x,\quad x={\rm o}(N), \quad x \rightarrow
\infty.
\]

The limits $x \rightarrow 0$ and $N \rightarrow \infty$ do not
commute; the evaluation of the asymptotics of $\Ip(x)$ as $x
\rightarrow 0$ requires first the computation of the limit $x
\rightarrow 0$ and then of the limit $N\rightarrow \infty$. The
application of Szeg\H{o}'s theorem clearly prevents this,
therefore provides information only for relative large $x$.
However, Dyson's electrostatic model for the CUE ensemble leads
naturally to the assumptions that for small $x$, as $N \rightarrow
\infty$, the distance of the roots of $Z\,'(U,z)$ from the unit
circle is on average of order $1/N$ and is distributed like the
square of the spacings between phases of consecutive eigenvalues
of unitary matrices in the CUE ensemble (appropriately rescaled).
These two simple hypotheses give a conjecture for $\Ip(x)$ as $x
\rightarrow 0$ whose agreement with numerical experiments covers
with high accuracy the region where the large $x$ asymptotics
fails.

Unfortunately, the zeros of $\zeta\,'(s)$ are poorly understood,
and there is not even a conjecture for their horizontal
distribution to compare with the results derived in this paper.
Given that it seems extremely difficult to obtain an analytical
expression of such a quantity, we believe that it would be
interesting and worthwhile to conduct a thorough numerical study
as independent verification of the model presented here.

\ack

I would like to thank Brian Conrey, the director of the American
Institute of Mathematics, for numerous conversations on this
subject and for kind hospitality while this project was carried
out. I am also indebted for helpful discussions to Mark Dennis,
Persi Diaconis, David Farmer, John Hannay, Chris Hughes, Jon
Keating, Michael Rubinstein and Nina Snaith. This work was
supported by a Royal Society Dorothy Hodgkin Research Fellowship
and in part by National Science Foundation, grant n. 0074028.

\renewcommand{\theequation}{A\arabic{equation}}

\section*{Appendix. Derivatives of Szeg\H{o}'s strong asymptotic formula}
\label{Sfd}

In this appendix we show that in the cases considered in the
present paper, the error term in formula~\eref{Sf} remains small
when differentiated with respect to $\alpha$; in other words, we
want to prove that
\begin{eqnarray}
\label{swth}
\fl \frac{\partial^2}{\partial \alpha^2} \Biggl
[D_{N-1}[\exp(\rmi g)](\alpha) + D_{N-1}[\exp(\rmi h)](\alpha)
 \Biggr]_{\alpha=0}  \nonumber \\
 \sim  \frac{\partial^2}{\partial \alpha^2}
 \Biggl[\, \exp\left(- E(g) \right) + \exp\left(-E(h)\right)
 \Biggr]_{\alpha=0}, \quad N \rightarrow \infty,
\end{eqnarray}
where $E(g)$ and $E(h)$ are defined in equations (3.3)\footnote{To
simplify the notation and emphasize the dependence on $\alpha$, we
omit the variables $w$ and $z$ in this appendix.}. The main idea
of the proof is quite simple: we first represent the Toeplitz
determinants with exact formulae and differentiate them with
respect to $\alpha$; then we take the limit $N \rightarrow
\infty$. We shall consider only $D_{N-1}[\exp(\rmi g)]$, as the
proof for $D_{N-1}[\exp(\rmi h)]$ is completely analogous.

The identity~\eref{H-S_id} allows us to write
\[
D_{N-1}[\exp(\rmi g)](\alpha) = \int_{\UN} \exp\left(\rmi
\sum_{k=-\infty}^{\infty} \hat{g}_k(\alpha) \Tr (U^k)\right)\rmd
\mu (U).
\]
Replacing the exponential function by its power series gives
\[
\int_{\UN}\prod_{k=1}^{\infty}
\sum_{a_k=0}^{\infty}\frac{\left(\rmi \hat{g}_k(\alpha) \Tr
(U^k)\right)^{a_k}}{a_k!}
\sum_{b_k=0}^{\infty}\frac{\overline{\left(-\rmi \hat{g}_k(\alpha)
\,\Tr (U^k)\right)^{b_k}}}{b_k!} \, \rmd \mu(U).
\]
 Integrals over
$\UN$ of product of traces of unitary matrices have been computed
by Diaconis and Shahshahani~\cite{DS94}. Let $\lambda_j$ be
nonnegative integers such that
\[
\lambda_1 \ge \lambda_2 \ge \ldots \ge \lambda_s
\]
and
\[
L = \lambda_1 + \lambda_2 + \cdots + \lambda_s =  1a_1 + 2a_2 +
\cdots + r a_r,
\]
where $a_k$ denotes the number of times the integer $k$ appears
among the $\lambda_j\,$s. We call $\lambda_a = (\lambda_1,
\lambda_2, \ldots, \lambda_s)$ a partition of $L$.  Consider the
integral
\[
I(\lambda_a,\lambda_b)=\int_{\UN} \prod_{k=1}^r
\left(\Tr(U^k)\right)^{a_k}\overline{\left(\Tr(U^k)\right)^{b_k}}
\, \rmd \mu(U).
\]
It turns out that $I(\lambda_a,\lambda_b)=0$ unless
$\lambda_a=\lambda_b$; furthermore, we have (see,
e.g.,~\cite{BD02})
\begin{equation}
\label{cas}
 \cases{I(\lambda_a,\lambda_b)= \delta_{\lambda_a \lambda_b}
 \prod_{k=1}^r k^{a_k} a_k! & if $L \le N$, \\
I(\lambda_a,\lambda_b) \le \delta_{\lambda_a \lambda_b}
 \prod_{k=1}^r k^{a_k} a_k! & if $L > N$.}
\end{equation}
Therefore, we obtain
\begin{equation}
\label{persi}
 D_{N-1}[\exp(\rmi g)](\alpha) =
\sum_{\lambda_a}I(\lambda_a,\lambda_a) \prod_k
(-1)^{a_k}\frac{\abs{\hat{g}_k(\alpha)}^{\, 2a_k}}{(a_k!)^2},
\end{equation}
where the sum is over all partitions and the product over all the
integers (without multiplicity) of a given partition. Because of
equation \eref{cas}, asymptotically formula \eref{persi} tends to
\begin{eqnarray*}
\sum _{\lambda_a} \prod_k (-1)^{a_k} \frac{k^{a_k}
\abs{\hat{g}_k(\alpha)}^{\, 2a_k}}{a_k!}& = \prod_{k=1}^{\infty}
\sum_{a_k=0}^{\infty}\left(-1\right)^{a_k}
\frac{k^{a_k}\abs{\hat{g}_k(\alpha)}^{\,2a_k}}{a_k!}\\
& = \exp\left(- \sum_{k=1}^\infty k \abs{\hat{g}_k(\alpha)}^2
\right).
\end{eqnarray*}
This proof of the strong Szeg\H{o} limit theorem was derived by
Bump and Diaconis~\cite{BD02}.

In order to prove~\eref{swth} we need to take the second
derivative with respect to $\alpha$ of~\eref{persi} and then the
limit $N\rightarrow \infty$. Differentiating the right-hand side
of equation~\eref{persi} is tedious but elementary. We shall carry
out only the first derivative, since the second one is completely
analogous. We have
\begin{eqnarray}
\label{fder}
 \fl \frac{\partial D_{N-1}[\exp(\rmi
g)](\alpha)}{\partial \alpha} &= \sum_{\lambda_a}
I(\lambda_a,\lambda_a)\sum_k \left(-1\right)^{a_k} \frac{\partial
\abs{\hat{g}_k(\alpha)}^2}{\partial
\alpha}\frac{\abs{\hat{g}_k(\alpha)}^{\, 2\left(a_k-1\right)}}%
{\left(a_k-1\right)!a_k!}  \nonumber \\
& \quad \times \prod_{j \neq k}\left(-1\right)^{a_j}
\frac{\abs{\hat{g}_k(\alpha)}^{\, 2a_j}}{\left(a_j!\right)^2}.
\end{eqnarray}
In the limit $N \rightarrow \infty$ the right-hand side
of~\eref{fder} becomes
\[
-\sum_{k=1}^\infty k \, \frac{\partial
\abs{\hat{g}_k(\alpha)}^2}{\partial \alpha} \exp\left(-
\sum_{k=1}^\infty k \abs{\hat{g}_k(\alpha)}^2\right)=
-\frac{\partial E(g)}{\partial \alpha} \exp\left(-E(g)\right),
\]
which is the same expression obtained by differentiating
Szeg\H{o}'s strong asymptotic formula. Similarly,
differentiating~\eref{fder} and then taking the limit $N
\rightarrow \infty$ gives
\begin{eqnarray}
\label{2der}
\frac{\partial^2 D_{N-1}[\exp(\rmi g)](\alpha)}{\partial \alpha^2}
& \sim \left[\left(\frac{\partial E(g)}{\partial \alpha}\right)^2
- \frac{\partial^2 E(g)}{\partial \alpha^2} \right] \nonumber \\
& \quad \times \exp\left(-E(g) \right), \quad N \rightarrow \infty.
\end{eqnarray}
This completes the proof of~\eref{swth}.

\end{document}